\documentclass[aps,prd,
nofootinbib,
preprintnumbers,
twocolumn,
showpacs, 
floatfix,
superscriptaddress
]{revtex4}

\pdfoutput=1

\usepackage{amsmath}
\usepackage{amssymb}
\usepackage{epsfig}
\usepackage{graphicx}
\usepackage{multirow}
\usepackage{color}
\usepackage[normal]{subfigure}
\usepackage{rotating}
\usepackage{hyperref}

\begin{document}

\thispagestyle{empty}
\preprint{ULB-TH/14-07}

\title{Adiabatic contraction revisited: implications for primordial black holes}

\author{Fabio Capela}
\affiliation{Service de Physique Th\'{e}orique, 
Universit\'{e} Libre de Bruxelles (ULB),\\CP225 Boulevard du 
Triomphe, B-1050 Bruxelles, Belgium}

\author{Maxim Pshirkov~\footnote{Dynasty Foundation fellow}}
\affiliation{Sternberg Astronomical Institute, Lomonosov Moscow State University, 
Universitetsky prospekt 13, 119992, Moscow, Russia}

\affiliation{Pushchino Radio Astronomy Observatory, 
Astro Space Center, Lebedev Physical Institute  Russian Academy of Sciences,  
142290 Pushchino, Russia}

\affiliation{Institute for Nuclear Research of the Russian Academy of Sciences, 117312, 
Moscow, Russia}

\author{Peter Tinyakov}
\affiliation{Service de Physique Th\'{e}orique, 
Universit\'{e} Libre de Bruxelles (ULB),\\CP225 Boulevard du 
Triomphe, B-1050 Bruxelles, Belgium}


\begin{abstract} 
We simulate the adiabatic contraction of a dark matter (DM)
distribution during the process of the star formation, paying
particular attention to the phase space distribution of the DM
particles after the contraction. Assuming the initial uniform density
and Maxwellian distribution of DM velocities, we find that the number
$n(r)$ of DM particles within the radius $r$ scales like $n(r) \propto
r^{1.5}$, leading to the DM density profile $\rho\propto r^{-1.5}$, in
agreement with the Liouville theorem and previous numerical
studies. At the same time, the number of DM particles $\nu(r)$ with
periastra smaller than $r$ is parametrically larger, $\nu(r) \propto
r$, implying that many particles contributing at any given moment into
the density $\rho(r)$ at small $r$ have very elongated orbits and
spend most of their time at distances larger than $r$. This has
implications for the capture of DM by stars in the process of their
formation. As a concrete example we consider the case of primordial
black holes (PBH). We show that accounting for very eccentric orbits
boosts the amount of captured PBH by a factor of up to $2\times 10^3$
depending on the PBH mass, improving correspondingly the previously
derived constraints on the PBH abundance.

\end{abstract}

\maketitle

\section{Introduction}
\label{sec:introduction}

Astrophysical and cosmological observations have provided a compelling
evidence that about 28\% of the energy density of the Universe is contained in
the form of a non-relativistic non-baryonic dark matter (DM)
\cite{2013arXiv1303.5076P}. Despite the extensive experimental efforts, all the
attempts at direct and indirect non-gravitational detection of this matter
have been unsuccessful so far, so that the DM nature remains essentially unconstrained, leaving
room for a diversity of candidates. It is often assumed that the DM is
composed of some kind of new particles beyond the Standard Model --- axion-like
particles, sterile neutrinos, weakly interacting massive particles are the
most common examples. However, other candidates such as primordial black
holes (PBH) may provide a viable alternative. In the latter case, the
advantage is that no new particles beyond the Standard Model are required.

Apart from direct and indirect searches, competitive constraints on DM
properties may be obtained from observations of stars where the DM could have
been accumulated and produced observable effects. In the Sun, the DM annihilation
may result in the observable flux of high energy neutrinos \cite{2009JCAP...08..037N}.  DM may
also induce abnormal asteroseismological effects~\cite{Lopes:2012af} or
suppress convection zones~\cite{Casanellas:2012jp,Casanellas:2013nra} or even
modify the transportation properties throughout the
star~\cite{Frandsen:2010yj,Horowitz:2012jd}.

More catastrophic effects may result from the DM accumulation in compact stars
such as neutron stars (NSs) or white dwarfs(WDs). If the particles of DM are
not self-annihilating (e.g., asymmetric dark matter), the accumulated amount
of DM may become sufficient to form a black hole (BH) inside the compact
star~\cite{Kouvaris:2010jy,Kouvaris:2010vv,Kouvaris:2011fi}. Because of a much
higher density of nuclear matter in compact stars as compared to main sequence
stars, the accretion is sufficiently efficient to destroy the star in a short
time (see Ref.~\cite{Kouvaris:2013kra} for a detailed discussion, including
the role of the angular momentum). In this case mere observations of compact
stars imply constraints on the DM
properties~\cite{Kouvaris:2010jy,Kouvaris:2010vv,Kouvaris:2011fi}.

The same considerations obviously apply to the DM composed of
PBHs, in which case there is no need to
accumulate DM in order to form a BH. If even a single PBH is captured by a compact
star, the latter gets destroyed. Requiring that the probability of such an
event is small leads to the constraints on the PBH
abundance~\cite{Capela:2013yf,Capela:2012jz}.  

The key quantity which determines the strength of the constraints is the
amount of captured DM. There are two different capture mechanisms. 
A star can capture DM from its surrounding environment, such as the Galactic
halo, during its lifetime. The DM particles passing through the star may
interact with the nucleons, losing enough energy to become gravitationally
bound~\cite{1995PhRvD..51..328J,2009JCAP...08..037N}.  
Then each subsequent orbit will also pass through the
star, so that eventually, after many collisions, the DM particle will sink to
the center of the star. Such capture process can lead to the accumulation of
a considerable amount of DM inside the compact star throughout its
lifetime~\cite{Kouvaris:2010jy}.

The DM could also be captured during the star formation. A
main-sequence star is formed from the gravitational collapse of a
prestellar core in a giant molecular cloud. In the course of this
process, the DM that was initially gravitationally bound to the core
undergoes adiabatic contraction, forming a cuspy profile centered at
the star, with the density $\rho(r)$ behaving like $\rho(r)\propto
r^{-3/2}$. Some of this DM ends up captured inside the star.  This
mechanism was first discussed in~\cite{Sivertsson:2010zm} and more
recently in~\cite{Capela:2012jz} where the constraints on the
abundance of PBHs were derived. Note that the adiabatic contraction is
a purely gravitational phenomenon that assumes nothing about the
DM-to-nucleon cross section~\cite{Gnedin:2004cx}.

In this paper we study in more detail the adiabatic contraction of the DM
caused by the star formation process, with the final goal to obtain a more
precise estimate of the amount of the DM captured by a star as a result of its
formation. It turns out that for this purpose the (usually considered) DM
density profile after the contraction is not sufficient, and one needs to know
in more detail the distribution of DM in the phase space. As we will show,
the detailed calculations lead to the results that are qualitatively different
from the estimates based on the contracted DM density profile alone.  

To see what is the point, recall that in order to get captured, the DM
particles have to lose their energy by interactions with the star
material. Therefore, one needs to calculate the number of particles whose
orbits cross the star after the adiabatic contraction. In
Ref.~\cite{Capela:2012jz}, this number was estimated by taking the DM density
profile after the contraction and integrating it over the volume of the
newly-formed star. However, this estimate does not account correctly for
particles that spend only a small fraction of time inside the star, because
their contribution into the density is correspondingly suppressed. At the same
time these particles can still get captured because their orbits will cross
the star again and again, and eventually they will get captured if given
enough time. The question is how big is the number of such particles. 

To quantify this effect, we have performed a simulation of the adiabatic
contraction process where we have measured, in addition to the density at a
given distance, the number of particles that have orbits with the periastron
smaller than given radius $r$. Our key observation is that the number of such
particles scales differently at small $r$ than the number of particles that
{\em are} within $r$ at a given moment: while the latter behaves like $\propto
r^{1.5}$ (which corresponds to the density having a spike $\rho\propto
r^{-1.5}$), the former goes as $r$. This means that at small $r$ there are
much more particles that {\em ever pass} within $r$, than actually {\em are}
within $r$ at a given moment. If these particles have enough time to lose
their energy and get captured, this would substantially boost the amount of
captured DM and improve the constraints accordingly.  For relevant values of
parameters, the boost factor can be as large as $\sim 2\times 10^3$. The
estimate of Ref.~\cite{Capela:2012jz} based on the density profile is,
therefore, by far over-conservative.

When these considerations are applied to the case of PBH, the previously
derived constraints~\cite{Capela:2012jz} get improved. The improvement factor
depends on the PBH mass that determines the energy loss time and for some
masses can reach its maximum value of $\sim 2\times 10^3$.

The rest of this paper is organized as follows. In Sect. II we discuss the
capture of DM by adiabatic contraction during the star formation stage.  In
Sect. III we present the resulting constraints coming from the enhancement of
the DM capture in the case the latter is composed of PBHs. Finally, in Sect. IV
we present our conclusions.

\section{Adiabatic contraction of DM during star formation}
\label{sec:adiab-contr-dm}

In this section, we discuss the adiabatic contraction of DM during the process
of star formation, paying particular attention to the final distribution of
particle orbits.

\subsection{Star formation and adiabatic contraction of DM}
\label{sec:adiab-contr}

Stars are formed in giant molecular clouds~(GMCs) with a typical mass of
$10^5-10^6M_\odot$ and density $\rho_B \sim 500$~GeV/cm$^3$. Such GMCs are
initially supported against gravitational collapse by several pressure
components ~\cite{1991ApJ...371..296M,1987ARA&amp;A..25...23S}. When this
support gets reduced due to dissipation, GMCs fragment into smaller clumps,
each forming a protostar after the contraction of the gas. The newly formed
protostars have a central blob and a surrouding disk of gas. Once the central
part has accumulated most of its main-sequence mass from the surrounding disk
of gas, it is considered to be a pre-main-sequence star. Note that such
process takes much longer than the gravitational free-fall time
$t_{\text{ff}}\sim (G \rho_0)^{-1/2}$, where $\rho_0$ is the average density
of the protostar cloud. The initial baryonic density is known from
observations of the SCUBA instrument~\cite{Kirk:2005ng} and fits well a
Bonnor-Ebert profile~\cite{Kirk:2005ng} that can be approximated by a uniform
density profile in the core. At the initial stages, such baryonic density is
quite high, so for most environments the gravitational effect of the DM on the
formation of stars is negligible. For all practical purposes, the behavior of
DM is determined by the gravitational potential of the baryons.

The baryons contracting into a protostar create a time-dependent gravitational
potential which drags the DM particles along with the contracting baryons. As
a result, the DM distribution develops a density profile that is peaked at the
core of the prestellar cloud. Since this process takes much more time than the
free-fall time $t_{\text{ff}}$, certain quantities are adiabatically
conserved, which may be used in analytical estimates, provided some simplifying
assumptions are made about the particle orbits. In realistic cases, however,
one has to resort to numerical simulations.

\subsection{Simulation of DM orbits}
\label{sec:simulation-orbits}

The star formation stage relevant for this paper is the formation of the star
from a pre-stellar core. We adopt the core parameters following
Ref.~\cite{Capela:2012jz}, the typical values being the baryonic density $\rho
\sim 5\times 10^6$GeV/cm$^3$ and the size $\sim 5000$~AU. This baryonic
density is much larger than the DM density in any of the environments that
will be considered in this paper, so we neglect the DM contribution to the
gravitational potential. This simplifies the simulations enormously as the
DM trajectories may be evolved, one at a time, in the same time-dependent
gravitational potential created by the contracting baryons. 

In the adiabatic approximation it is not important how exactly the
gravitational potential changes --- only the initial and final states
matter. We have modeled the gravitational potential of the contracting baryons
by a two-component mass distribution: the uniform spherical cloud and a point
mass in its center. The total mass of the cloud and the value of the central
mass are chosen in such a way that the latter linearly grows from zero at
$t=0$ to the maximum value at $t=T$, while the former decreases from the
maximum value to zero, the total mass of the system being constant.  In order
to satisfy the adiabatic condition requiring the time of the star formation
process to take much longer than the free-fall time $t_{\text{ff}}$, this
change is performed slowly over a period of time $T$ that is at least several
times longer than $t_{\text{ff}}$.

For the initial DM distribution we take a uniform distribution in space
characterized by some background DM density $\bar \rho_{\rm DM}$, and the
Maxwellian distribution in velocities with the dispersion $\bar v$,
\begin{equation}
d n = \bar n_{\rm DM} \left({3\over 2\pi \bar v^2}\right)^{3/2}
\exp \left\{{-3v^2\over 2 \bar v^2}\right\} d^3v , 
\label{eq:Maxwell}
\end{equation}
where $\bar n_{\rm DM} = \rho_{\rm DM}/m_{\rm DM}$ is the mean DM number
density. The assumption of the Maxwellian distribution in velocities is not an
essential one, because only the particles that are gravitationally bound to
the pre-stellar core are affected by the contraction of baryons. These
particles have very small velocities as compared to $\bar v$ even for $\bar v$
as low as a few km/s. Thus, any distribution which is flat around $v\sim 0$
will produce the same results. 

An estimate of the initial density of DM that is {\em gravitationally bound}
inside the pre-stellar core has been computed for several star masses
in~\cite{Capela:2012jz}. This estimate is essentially based on the assumption
that the formation of the pre-stellar core does not change substantially the
DM phase space density (the Liouville's theorem) as compared to the ambient DM
distribution. From eq.~(\ref{eq:Maxwell}), at zero velocity $v$ the ambient
phase space density is (cf. ~\cite{Tremaine:1979we})
\begin{equation}
\mathcal{Q}_{\text{max}}=\left(\frac{3}{2\pi} \right)^{3/2} 
\frac{\bar{\rho}_{\text{DM}}}{m_{\text{DM}}^4 \bar{v}^3}.
\end{equation}
So, the density of DM gravitationally bound to the pre-stellar core is
suppressed as compared to the ambient DM density by the factor $(v /
\bar{v})^3\ll 1$, $v$ being the escape velocity from the pre-stellar
core. This suppression factor is included in our estimates.

The simulation proceeds as follows. At $t=0$ we inject a DM particle with a
random uniformly distributed initial position and velocity (keeping in mind
that at $v\ll \bar v$ the Maxwellian distribution reduces to a uniform
one). If the particle is gravitationally bound at $t=0$, it is evolved through
the equations of motion, otherwise it is rejected. The basic setup is,
therefore, the same as in Ref.~\cite{Capela:2012jz}. We have simulated about
$3\times 10^7$ trajectories in this way.

The density profile can be obtained by recording the positions of particles at
a randomly chosen time $t>T$ (i.e., when the baryonic contraction is
finished). Their distribution in $r$ can be then converted into particle
density. This information, however, is not enough for an accurate estimate of
the number of captured particles because many trajectories pass through the
star, but spend most of their time outside. The number of such trajectories is
not reflected in $n(r)$, and accounting for them is the main goal of the current
calculation (and the main difference with Ref.~\cite{Capela:2012jz}).  So, to
have a complete information about each trajectory we calculate and record the
periastron $r_{\text{min}}$ and apastron $r_{\text{max}}$ of each particle
orbit at the end of star formation (these, of course, do not depend on time 
once the contraction is finished).
The distribution of $r_{\text{min}}$ determines the number of particles that
ever get within a given distance from the center.  

In order to calculate the density profile after the contraction 
one has to sample the particle position $r$ uniformly in time. 
The actual simulation of the trajectory after the
contraction is not needed for this purpose: instead one can use the Kepler's
second law, which states that an imaginary line connecting the particle's
position to the center of the star sweeps out the same area in the same amount
of time. Therefore, for the trajectory defined by the given values of
$r_{\text{max}}$ and $r_{\text{min}}$ the quantity
\begin{eqnarray}
\xi(x)&=&\frac{1}{2}-\frac{2}{\pi(1+\mathcal{R})}
\sqrt{(\mathcal{R}-x)(x-1)} \nonumber \\
&-&\frac{1}{\pi}\arctan
\left(\frac{1+\mathcal{R}-2x}{2\sqrt{(\mathcal{R}-x)(x-1)}} \right)
\label{eq:area}
\end{eqnarray}
where $x=r/r_{\text{min}}$, $\mathcal{R}=r_{\text{max}}/r_{\text{min}}$ and
$\xi=A/A_{\text{tot}}$ with $A_{\text{tot}}$ half of the total area of the
orbit, is proportional to the time.  This function is constructed in such a
way that $\xi(1)=0$ and $\xi(\mathcal{R})=1$. Thus, one can generate a random
uniformly distributed $\xi \in [0,1]$ and solve eq.~(\ref{eq:area}) for $x$,
obtaining $r = x r_{\rm min}$ which is distributed in a way that the uniform
time sampling would give. This is computationally efficient, and can be done
several times per each simulated trajectory. We have checked that the
resulting distribution of $r$ agrees with the directly measured one,
Ref~\cite{Capela:2012jz}.

\begin{figure}
\begin{picture}(220,150)
\put(-10,0){\includegraphics[width=0.98\columnwidth]{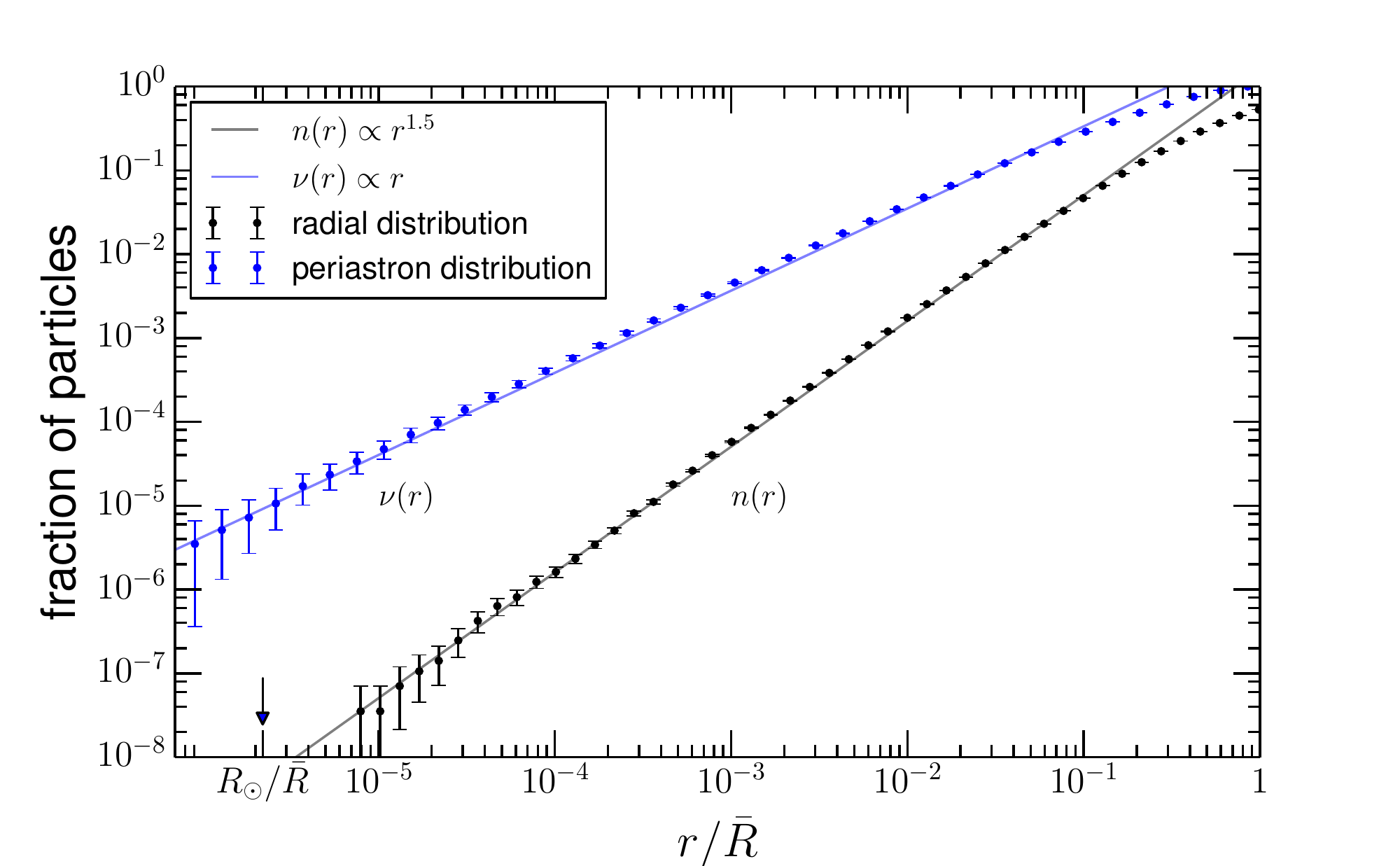}}
\end{picture}
\caption[width=0.95\columnwidth]{
{\it Lower curve:} The fraction of particles $n(r)$ 
that are found within radius $r$ at the
end of the adiabatic contraction. $\bar R$ is the
initial radius of the prestellar core. {\it Upper curve:} The fraction of particles
$\nu(r)$ whose orbits have the periastron smaller than $r$. Lines show the 
power laws $n(r) \propto r^{1.5}$ and $\nu(r) \propto r$. The errorbars  
represent statistical errors. }
\label{fig:distribution}
\end{figure}
Fig.~\ref{fig:distribution} shows the resulting distributions as a function of
$r/\bar R$, where $\bar R$ is the initial radius of the cloud. The actual star
radius is also indicated on the plot. The lower curve shows the fraction
$n(r)$ of particles that are, in a given moment of time, within the radius $r$
from the center. Points with errors represent the results of the
simulation. The straight line is a fit by the power law $n(r) \propto
r^{1.5}$. Bearing in mind that $n(r)$ is proportional to the density
integrated up to the radius $r$, we conclude that $\rho(r) \propto r^{-1.5}$,
in agreement with the results of Refs.~\cite{Capela:2012jz} and
\cite{Ullio:2001aa}\footnote{We would like to thank the anonymous referee for
  pointing out the last reference.}

The upper curve shows the fraction $\nu(r)$ of particles whose orbits have
periastra within the radius $r$. One can see that the fraction of such
orbits is larger than $n(r)$. The reason is that, after the adiabatic
contraction, a considerable amount of DM particles have very elongated orbits
and spend most of their time outside the region of radius $r$. Indeeed,
$\nu(r)$ scales differently,  $\nu(r) \propto r$.  The ratio of the two curves 
gives the enhancement factor as a function of the radius. At the star radius
$r=R_\odot$ this factor is $\nu(R_\odot) /n(R_\odot) = 1.84\times 10^3$. This
is our key observation and the source of differences with
Ref.~\cite{Capela:2012jz}. 

A few comments are in order. First, the number of simulated trajectories is
sufficient to directly measure $\nu(r)$ and almost sufficient to measure
$n(r)$ -- the required extrapolation in the latter case is by a factor of a
few only. Second, we have checked that the results do not depend on the
duration $T$ of the baryon contraction as long as $T\gtrsim
3t_{\text{ff}}$. This means that the adiabatic approximation is a good
approximation as long as $T\gtrsim 3 t_{\text{ff}}$. Remark that the
natural timescale of the star formation is~$T \gg
t_{\text{ff}}$~\cite{2009ApJ...705..650W}.  Finally, we would like to point
out that our results are in agreement with the Liouville's theorem which 
requires the phase space density of DM particles to stay constant during the
star formation process. This implies that the density of DM after the
contraction cannot be more cuspy than $\rho\propto r^{-1.5}$, which is
satisfied in our simulations. The distribution in periastra, on the
contrary, has no straightforward limitations from the Liouville's theorem.

To summarize, as a result of the formation of a star, the DM
bound to the prestellar core experiences the adiabatic contraction and
develops a cuspy profile with the density $\rho\propto r^{-1.5}$. Moreover,
the distribution of the DM particles in orbits becomes such that there are
much more particles that {\em ever} come within given radius $r$ than there
 {\em are} with $r$ at any given moment. For $r=R_\odot$ the enhancement
factor is $\sim 2\times 10^3$. Whether all these DM particles can be finally
captured by the star depends on whether there is enough time for them
to lose energy via the DM-nucleon interactions.  
As a concrete example, in the next section we work out the implications of 
this enhancement factor for the case of PBHs as DM candidates.

\section{Constraints on PBH}
\label{sec:constraints-pbh}

\subsection{Existing constraints}
\label{sec:existing-constraints}

For completeness, we start with the brief summary of the already existing
constraints. Depending on the mass of PBHs, different effects result in
constraints on their fraction. First, PBHs with masses $m_{\text{BH}}\leq
5\times 10^{14}\text{g}$ evaporate through Hawking's radiation in a time
shorter than the age of the Universe, making them unviable DM
candidate. At higher masses, although PBHs are able to survive until present
epoch, they evaporate too efficiently and overproduce the observed $\gamma$-ray
background unless their mass is larger than $\mbox{(a few)} \times
10^{16}$~g~\cite{Carr:2009jm,Sreekumar:1997un,1998ApJ...499L...9A}.

For PBHs with masses $m_{\text{BH}}\geq 10^{17}\text{g}$ the Hawking radiation
is negligible.  Still, constraints can be imposed by looking at microlensing
effects that such PBHs induce when passing in front of a bright source. In
this way, the EROS microlensing survey and the MACHO collaboration derived
constraints on the fraction of PBHs in the Galactic halo for the range of
masses $10^{26}~\text{g}< m_{\text{BH}}<
10^{33}~\text{g}$~\cite{Tisserand:2006zx,1998ApJ...499L...9A} that may be
improved in the near future~\cite{2013ApJ...767..145C,2013arXiv1307.5798G}.
In the same region of masses, the effect of superradiant instabilities of
rotating PBHs on the cosmic microwave background has been studied
in~\cite{2013PhRvD..88d1301P}, leading to interesting constraints.

For $m_{\text{BH}}>10^{33}~\text{g}$, it has been found
in~\cite{Ricotti:2007au} that the accretion of gas by PBHs before
the non-linear stage of the large scale structure formation leads 
to emission of X-rays that induces changes in the
cosmological parameter estimates. Using both data from the three-year
Wilkinson Microwave Anisotropy Probe (WMAP3) and the COBE Far Infrared
Absolute Spectrophotometer, very stringent constraints have been derived
in~\cite{Ricotti:2007au}.

In the remaining mass window
$10^{17}~\text{g}<m_{\text{BH}}<10^{26}~\text{g}$, constraints from
femtolensing for masses $10^{18}~\text{g}<m_{\text{BH}}<10^{20}~\text{g}$ have
been claimed in Ref.~\cite{Barnacka:2012bm}. Finally, considering both the PBH
capture at the stage of star formation and during the lifetime, the
constraints have been derived in Refs.~\cite{Capela:2012jz,Capela:2013yf} in
the range of masses $10^{16}~\text{g}<m_{\text{BH}}<10^{24}~\text{g}$ assuming
the existence of dense DM cores in globular clusters at the epoch of star formation
as would be the case if
the latter were of a primordial origin. Note that much more stringent
constraints based on the PBH capture by neutron stars have been claimed
recently in Ref.~\cite{2014arXiv1401.3025P}. Such constraints
have been independently analyzed and criticized in~\cite{2014arXiv1402.4671C,Defillon:2014wla}.
The existing constraints are summarized in Fig.~\ref{fig:constraints_total}.

\subsection{Revised constraints on PBH from star formation}
\label{sec:revis-constr-pbh}

The complete information about the DM particle trajectories after the
adiabatic contraction allows one to make a more accurate estimate of
the number of captured PBHs than it was done in
Ref.~\cite{Capela:2012jz}. Since only the PBHs that can lose energy
can be captured, only the trajectories that pass through or in the
near vicinity of the star are relevant. To calculate the number of
captured PBHs, for each PBH trajectory of this type one has to
determine whether there is enough time for the PBH to lose energy and
end up in the compact remnant of the star --- WD or NS. 

Let us sketch the calculation of the capture time. Two stages of the
energy loss should be distinguished. At the first stage, the PBH
spends part of the time outside the progenitor star. Such orbits can
be approximately considered as radial, except a small number of
cases when the apastron is of order $R_*$. In this
approximation, the capture time has been estimated
in~\cite{Capela:2013yf} to be of order
\begin{equation}
t_{\text{capt}}\simeq 2 \tau 
\sqrt{\xi_0}\sim2\times 10^{8}\;\text{yrs} 
\left(\frac{10^{22}~\text{g}}{m_{\text{BH}}} \right), 
\label{eq:estimation_time_capture}
\end{equation}
where 
\begin{equation}
  \tau = \frac{\pi R_*^{5/2} v^2}{4Gm_{\text{BH}}\sqrt{GM} 
\ln \Lambda} \nonumber ,
\end{equation}
with $\xi_0=r_\text{max}/R_*$, $v$ the escape velocity of the star,
$\ln \Lambda$ the Coulomb logarithm~\cite{1987gady.book.....B} that
takes a value close to $\ln \Lambda\simeq 30$ for a main-sequence
star, and $R_*$ and $M$ the radius and the mass of the star,
respectively.  For the numerical value in
eq.~(\ref{eq:estimation_time_capture}) we have taken
$r_{\text{max}}/\bar{R}\simeq 0.1$ corresponding to a value of
$\xi_0\simeq 4.4\times 10^4$. However, when computing the constraints presented below
we have used the values of $r_\text{max}$ corresponding to simulated
trajectories, which is the main difference with the estimates of
Ref.~\cite{Capela:2012jz}. In this way we have determined, for each
PBH mass, the fraction of simulated PBH orbits that would be captured
by the star during its lifetime.

While deriving the constraints, we have also taken into
account the PBH orbits that do not cross the star, but come close enough to
lose energy through tidal interactions 
\cite{1977ApJ...213..183P}. However, the amount of PBHs captured
in this way turned out to be negligible.

The second stage starts when PBH is fully inside the star. It then
continues to lose energy through the dynamical
friction~\cite{1949RvMP...21..383C} and gradually sinks towards the
star center until the moment when the star turns into a compact
remnant. If the radius to which the PBH has been able to sink to
during the lifetime of the star is smaller than the radius of the
compact remnant, the latter inherits a PBH. The efficiency of the
dynamical friction grows with the PBH mass. At large masses all the
PBH captured by a star have time to sink to within the radius of the
future remnant in a lifetime of the star. At small masses only a
fraction of captured PBH can make it.  At the second stage there is no
difference with the calculations of
Ref.~\cite{Capela:2012jz}. Combining the two stages gives the fraction
of the PBHs that ends up inside the compact remnant.

The final step in deriving the constraints consists in relating the number of
simulated trajectories with the mean density of DM in a given
environment. This can be done as described in Ref.~\cite{Capela:2012jz} by
calculating the fraction of the DM particles which are gravitationally bound
to the prestellar core before the adiabatic contraction.  An important point
to keep in mind is that this fraction is proportional to the total DM density
and inversely proportional to the cube of the DM velocity dispersion.  Making
use of this scaling, the results for one particular DM density and velocity
dispersion can be rescaled to other values of these parameters.

To summarize, the difference between the present calculation and
Ref.~\cite{Capela:2012jz} is that there it was assumed that only PBHs
that are within the star at a given moment after the end of the
adiabatic contraction are captured by the star and may end up inside
the compact remnant, provided they sink sufficiently close to the
center by the time the remnant is formed. On the contrary, in the
present calculation all PBH trajectories that ever pass through the
star are taken into account. Whether there is sufficient time for a
PBH to be captured by the star is calculated individually for each
trajectory. The distribution of PBH over parameters of the orbits is
taken directly from the simulation.

\begin{figure}
\begin{picture}(220,150)
\put(-10,0){\includegraphics[width=0.98\columnwidth]{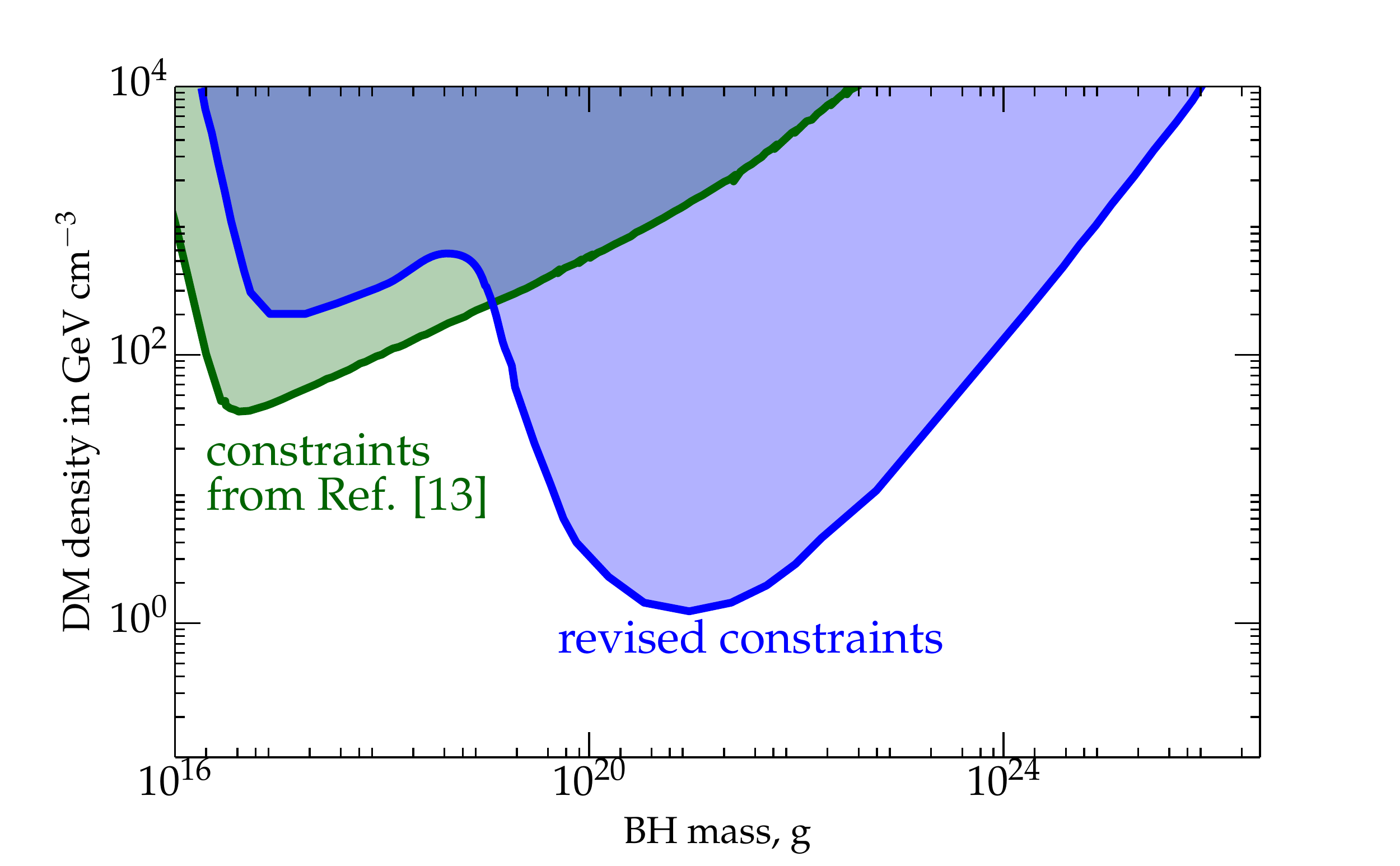}}
\end{picture}
\caption[width=0.95\columnwidth]{Constraints on the abundance of PBHs 
assuming the DM velocity dispersion of $7$~km/s. The
constraints derived in Ref.~\cite{Capela:2012jz} are in green, while the
revised ones are in blue. }
\label{fig:constraints_compare}
\end{figure}
The resulting constraints are shown in Fig.~\ref{fig:constraints_compare}.  To
obtain these constraints we have considered stars with masses $M_{\odot}\leq M
\leq 7M_{\odot}$, the progenitors of WDs, and stars with $8M_{\odot}\leq M
\leq 15 M_{\odot}$ which become NSs. WDs lead to better constraints for low
PBHs masses, while NSs are more competitive at high masses. The transition
between the two regimes is around $m_{\text{BH}}\sim 10^{20}~\text{g}$.  For
comparison, the constraints of Ref.~\cite{Capela:2012jz} are also presented in
Fig.~\ref{fig:constraints_compare}. Both constraints are derived assuming the
velocity dispersion of $\bar v = 7$~km/s. Such velocities are characteristic
of globular clusters and dwarf spheroidal galaxies. The constraints for other
velocity dispersions can be obtained by a simple rescaling as
described above.

At large PBH masses, the revised constraints are substantially more stringent
than those of Ref.~\cite{Capela:2012jz}. In this mass range the energy loss
mechanism is so efficient that all the gravitationally bound PBH that ever
cross the star have time to be captured and transferred to a compact
remnant.  As pointed out in Sect.~\ref{sec:adiab-contr-dm}, the number of such
trajectories turned out to be much larger than it was assumed in
Ref.~\cite{Capela:2012jz}. In this range of masses the gain in the total
captured mass (and therefore, the improvement in the constraints) is the ratio
between the two curves of Fig.~\ref{fig:distribution} calculated at $r=R_*\sim
R_\odot$, which is close to $1.8\times 10^3$. For stellar masses $M_*>8 M_{\odot}$, 
this number is reduced by a factor two.

At small PBH masses, the revised constraints are, on the contrary, somewhat 
worse than the estimate of Ref.~\cite{Capela:2012jz}. The reason for this is 
that some of the PBH that are inside the star by the end of the adiabatic
contraction, are in fact on elongated orbits and spend most of the time
outside the star. When the energy losses are not efficient, these PBH do not
have enough time to lose their energy and get captured by the
star. Such orbits were, but should not have been, included in the estimates in 
Ref.~\cite{Capela:2012jz}, hence the difference.

\begin{figure}
\begin{picture}(220,150)
\put(-10,0){\includegraphics[width=0.98\columnwidth]{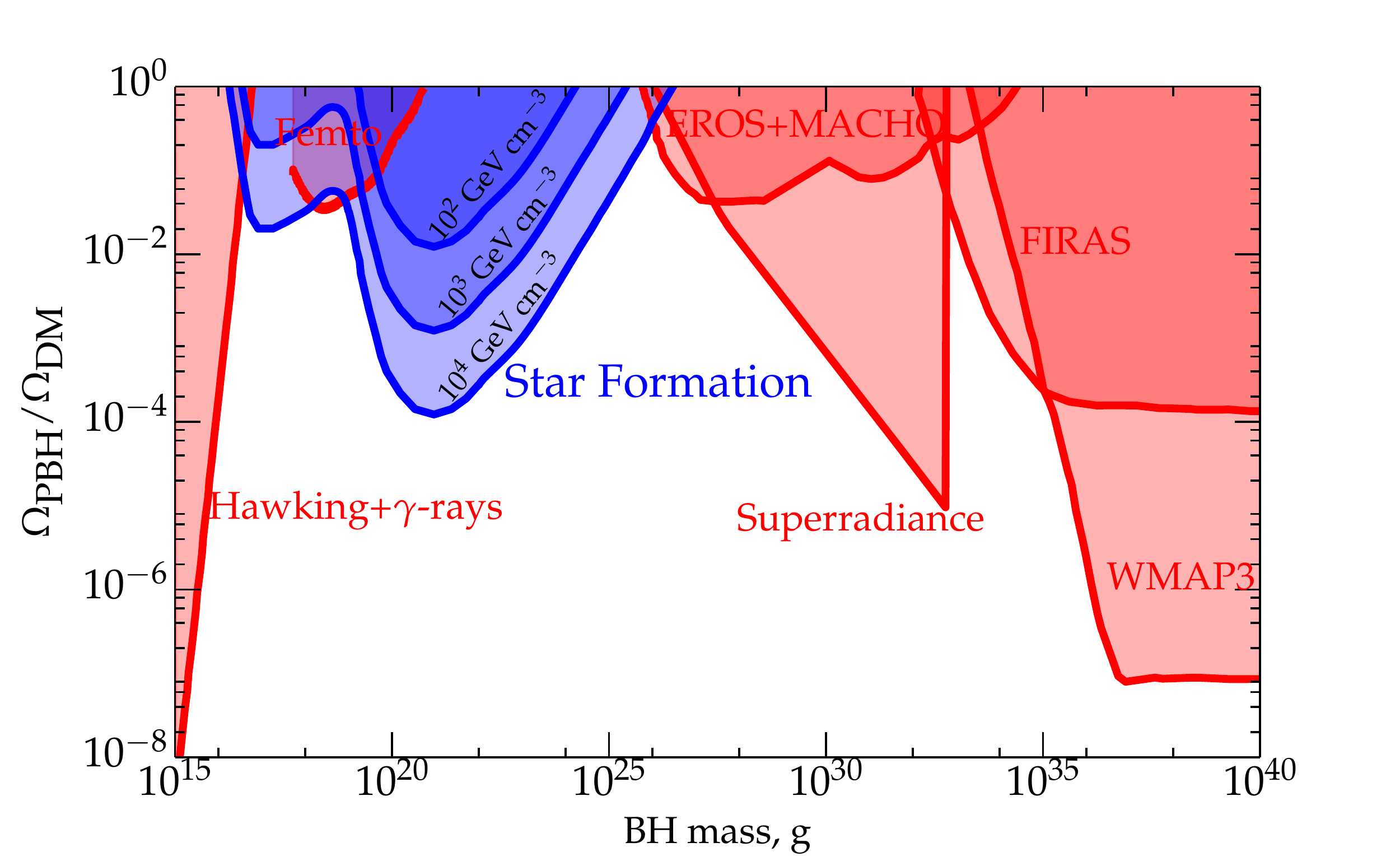}}
\end{picture}
\caption[width=0.95\columnwidth]{Constraints on the fraction of PBHs as DM.
  Shaded regions are excluded. The blue shaded regions correspond to the
  revised constraints derived in this paper assuming the DM densities of
  ($10^4$,$10^3$,$10^2$)~GeV/cm$^3$ and the velocity dispersion of $7$~km/s. The red shaded
  regions represent the existing constraints coming from various observations
  as explained in Sect.~\ref{sec:existing-constraints}.}
\label{fig:constraints_total}
\end{figure}
Clearly, most stringent constraints come from observations of compact stars in
regions with a high DM density and a low DM velocity dispersion.  As a
benchmark, we consider the values $\rho_{\text{DM}}= (10^{4},10^3,10^2)~\text{GeV}
~\text{cm}^{-3}$ and $\bar{v}=7~\text{km}~\text{s}^{-1}$. The constraints that
would result in these cases are shown in Fig.~\ref{fig:constraints_total}
together with the other existing constraints. The strongest constraints shown in light blue 
correspond to the highest value of DM density considered, i.e 
$\rho_{\text{DM}}\sim 10^{4}~\text{GeV}~\text{cm}^{-3}$ and decrease linearly for lower 
values of DM density. The corresponding conditions
may be present in the cores of metal-poor globular clusters at the epoch of star formation, 
if they are proved to be of a primordial origin~\cite{Bertone:2007ae} (see detailed
discussions in~\cite{Capela:2012jz,Capela:2013yf}). Another place where
similar conditions, albeit with somewhat lower densities, 
could exist are dwarf spheroidal galaxies that are
considered to be DM dominated~\cite{Strigari2008,2007PhRvD..75h3526S} and have
very low velocity dispersions~\cite{Strigari2008}. However, at present compact
objects such as NS or WD have not been observed in dwarf spheroidal
galaxies. Nonetheless, surveys for pulsars and X-ray
binaries have already revealed some glimpses of NSs existence in dSph
galaxies~\cite{2013IAUS..291..111R,Maccarone2005}, even though a clear
observation is still lacking.

\section{Conclusions}
\label{sec:conclusions}

In this paper we have considered the adiabatic contraction of DM during
the formation of a star. By simulating the behavior of $\sim 30$ million
particles, we reconstructed the phase space distribution of the DM at the end
of the star formation.  In particular, the number of particles $n(r)$ within a
given radius $r$ was found to be proportional to $r^{1.5}$, which corresponds
to the DM density profile $\rho(r) \propto r^{-1.5}$, in agreement with
previous calculations and the Liouville theorem.

At the same time, we have found that the adiabatic contraction creates a
rather special distribution of particle orbits. Namely, if one considers the
particles that contribute to $n(r)$ for a small $r$, a substantial ($O(1)$)
fraction of them have very elongated orbits with periastra smaller than
$r$. In fact, the number of particles $\nu(r)$ that have periastra smaller
than $r$ scales as $\nu(r) \propto r$. Such particles spend only a small
fraction of time close to the center, so their individual contributions to the
density at small $r$ are suppressed. However, they are numerous enough to
contribute non-negligibly to the density. At $r=R_\odot$, there are about
$1.8\times 10^3$ more particles that have periastra smaller than $r$ than
there are particles within $r$. 

This has implications for the DM capture by stars after their formation. A
large number of particles that constitute the DM cusp around the newly-formed
star have orbits that cross the star, which potentially leads to their
capture. This factor has not been taken into account in the previous estimates. 

As an application, we have considered the capture of PBH by stars, which leads
to the constraints on the PBH abundance. We have recalculated the constraints
of Ref.~\cite{Capela:2012jz} computing capture times individually for each
trajectory, and taking trajectories directly from the simulation of the
adiabatic contraction. In the range of PBH masses $10^{20}~\text{g}\lesssim
m_{\text{BH}} \lesssim 10^{26}~\text{g}$, the constraints are improved by
about 3 orders of magnitude due to the enhancement factor discussed above.
The resulting constraints are shown in Fig.~\ref{fig:constraints_compare}. 

The most stringent constraints are obtained from observations of compact stars
in the regions with high DM density and small velocity dispersion. The
examples corresponding to the densities $\rho =
(10^{4},10^3,10^2)~\text{GeV}~\text{cm}^{-3}$ and a low velocity dispersion
$\bar{v} = 7~\text{km}~\text{s}^{-1}$ are shown in
Fig.~\ref{fig:constraints_total}. Such conditions could have been present at
the cores of metal-poor globular clusters at the epoch of star formation if
they are of a primordial origin~\cite{Bertone:2007ae}, or --- with a DM density
somewhat smaller than $10^3~\text{GeV}/\text{cm}^3$ --- in dwarf spheroidal
galaxies~\cite{Strigari2008,2007PhRvD..75h3526S,2012MNRAS.420.2034S}. Even
though clear observations of compact objects in dwarf spheroidals are lacking
at the moment, first glimpses of NSs existence may have already been
observed~\cite{2013IAUS..291..111R,Maccarone2005}.

\acknowledgments
The authors are indebted to Michel Tytgat and Malcolm Fairbairn 
for valuable discussions and comments. 
M.P. acknowledges the hospitality of the Service de Physique Th\'{e}orique of
ULB where this work was initiated.  The work of F.C. and P.T. is supported in
part by the IISN and the Belgian Science Policy Belgian Science Policy under
IUAP VII/37. The work of M.P. is supported by RSF grant
No.~14-12-00146.

\bibliography{refs}
\bibliographystyle{apsrev}

\end{document}